\documentclass[journal=nalefd,manuscript=letter]{achemso}

\usepackage[utf8]{inputenc}
\usepackage{graphicx}
\usepackage{amsmath,amsfonts,amssymb}

\title{Structural Manipulation of Spin Excitations in a Molecular Junction}

\author{Maximilian Kögler}
\affiliation{Institut für Physik, Technische Universität Ilmenau, D-98693 Ilmenau, Germany}
\email{max.koegler@tu-ilmenau.de}

\author{Nicolas Néel}
\affiliation{Institut für Physik, Technische Universität Ilmenau, D-98693 Ilmenau, Germany}

\author{Laurent Limot}
\affiliation{Institut de Physique et Chimie des Matériaux de Strasbourg, Université de Strasbourg, F-67000 Strasbourg, France}

\author{Jörg Kröger}
\affiliation{Institut für Physik, Technische Universität Ilmenau, D-98693 Ilmenau, Germany}

\begin{document}

\maketitle

\begin{abstract}
Single metallocene molecules act as sensitive spin detectors when decorating the probe of a scanning tunneling microscope (STM)\@.
However, the impact of the atomic-scale electrode details on the molecular spin state has remained elusive to date.
Here, a nickelocene (Nc) STM junction is manipulated in an atom-wise manner showing clearly the dependence of the spin excitation spectrum on the anchoring of Nc to Cu(111), a Cu monomer and trimer.
Moreover, while the spin state of the same Nc tip is a triplet with tunable spin excitation energies upon contacting the surface, it transitions to a Kondo-screened doublet on a Cu atom.
Notably, the non-trivial magnetic exchange interaction of the molecular spin with the electron continuum of the substrate determines the spectral line shape of the spin excitations.
\begin{tocentry}
\centering
\includegraphics[width=\linewidth]{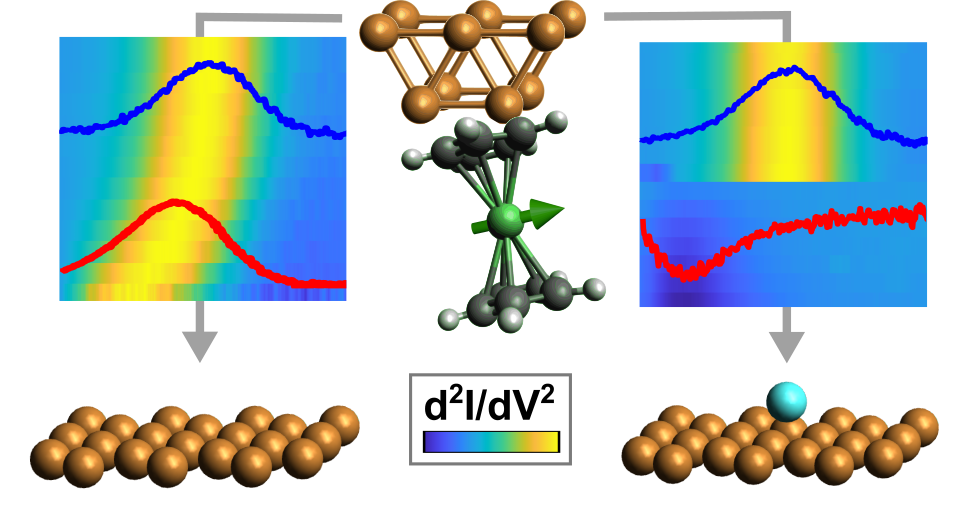}
\label{TOC}
\end{tocentry}
\end{abstract}


Single magnetic molecules are promising for a variety of applications in, e.\,g., quantum storage \cite{science_300_1130,jacs_129_2754}, quantum computing \cite{natchem_11_301,acsCS_1_488,advmat_35_2107534}, molecular spintronics \cite{natmat_7_179,nature_417_725,natrevmater_1_16044,natcommun_8_642,natcommun_9_3904}, and spin sensors \cite{science_306_466,science_312_1021,natmater_13_782,prl_115_237202,science_350_417}, because they often exhibit high spin-orbit coupling that provides magnetic anisotropy, which in turn gives rise to preferred spin directions without an external magnetic field. 
Moreover, molecular magnetism can be tailored by, e.\,g., hybridization with the environment \cite{natcommun_8_1974,prb_101_075414,natnanotechnol_9_64,nl_19_3288}, charge transfer \cite{nature_417_725,nl_17_1877,nl_18_88}, or structural relaxations caused by mechanical stress \cite{science_328_1370,nature_9_765,nl_15_4024}.
Apart from these appealing aspects from a nanotechnological perspective, the coupling of spin excitations of a single magnetic impurity to the electron continuum of a substrate represents an important model system for the understanding of fundamental topics in quantum physics, such as spin-polarized electron transport \cite{prl_61_2472,prb_39_4828,rpp_66_523,prl_101_116602,rmp_81_1495,prl_102_086805,natnanotechnol_6_185,njp_13_085011,prl_110_037202,nl_16_1450}, magnetic exchange interaction \cite{nature_446_522,nl_17_5660,prb_97_100401,natcommun_11_1197}, spin textures at surfaces \cite{science_323_915,nature_465_901,natphys_7_713,science_335_196,science_341_636}, and electron correlations \cite{jpcm_21_053001,prl_98_016801,pss_92_83,natcommun_11_6112,nl_23_8988,natphys_20_28}.

The scanning tunneling microscope (STM) appears to be particularly suitable for exploring quantum excitations in controlled environments, a favorable opportunity that results from the combined capabilities of imaging and manipulating matter at the atomic scale together with highly resolved spectroscopy \cite{natrevphys_1_703}.
Furthermore, the intentional termination of the tip apex with single atoms or molecules extends the control of the junction geometry and enables additional functionality, such as spin sensitivity.

\begin{figure}
\centering
\includegraphics[width=\textwidth]{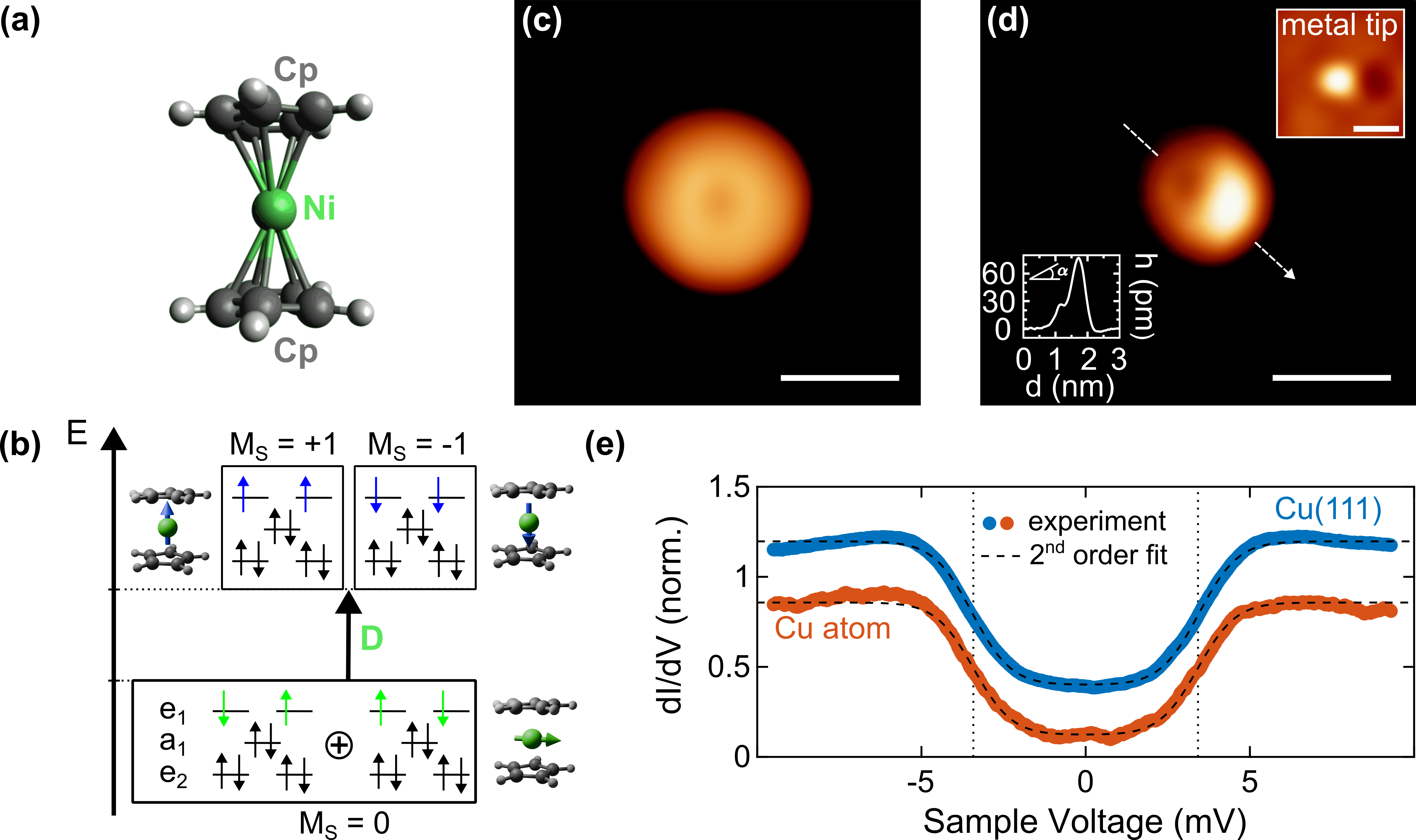}
\caption{Spin excitations of Nc.
(a) $D_{5h}$ symmetry of free-Nc molecular backbone with two parallel Cp planes sandwiching the Ni ion.
(b) Ligand-field-induced lifting of the Ni $d$-orbital degeneracy resulting in the ground state ($M_S=0$; 4 electrons in the low-energy $e_2$-terms ($d_{x^2-y^2}$, $d_{xy}$); 2 electrons on the middle-energy $a_1$-term ($d_{z^2}$); 1 electron in the $e_1$-terms ($d_{xz}$, $d_{yz}$) each) and the excited state ($M_S=\pm 1$), which are separated by the spin excitation energy $D>0$\@.
(c) Metal-tip STM image of Nc residing at a surface defect of Cu(111) (sample voltage: $-40\,\text{mV}$, tunneling current: $20\,\text{pA}$) revealing the parallel orientation of the top Cp plane to Cu(111)\@.
(d) STM image of a single Cu adatom recorded with an Nc tip ($-200\,\text{mV}$, $20\,\text{pA}$)\@. 
Insets: metal-tip STM image of the adatom (top) and cross-sectional profile acquired along the dashed line in (d) (bottom)\@.
Scale bars in all STM images indicate $1\,\text{nm}$.
(e) Spectra of $\text{d}I/\text{d}V$ acquired with an Nc tip atop Cu(111) (top) and a Cu adatom (bottom)\@.
Feedback loop parameters: $-10\,\text{mV}$ and $50\,\text{pA}$ (surface), $20\,\text{pA}$ (adatom)\@.
The spectra are normalized to their respective maximum and are vertically offset.
Dashed lines are fits to the data using a second-order dynamical exchange scattering model (see text)\@.
Dotted lines mark the spin excitation flanks at voltages $\pm D/\text{e}$ with $D$ resulting from the second-order fit.}\label{fig1}
\end{figure}

Very recently, nickelocene (NiCp$_2$, Cp: cyclopentadienyl, Fig.\,\ref{fig1}a) has been shown to retain its gas-phase spin triplet (spin quantum number $S=1$) ground state upon adsorption on surfaces as well as on the apex of an STM tip and to act as a molecular spin detector \cite{nl_17_1877,science_364_670,science_366_623}.
The frontier orbitals of Nc are mostly composed of Ni $d$-orbitals with admixture of C $p$-states.
The spin-orbit interaction induces
a uniaxial magnetic anisotropy $\mathcal{D}>0$ with an easy plane parallel to the Cp moieties \cite{nl_17_1877}.
The spin excitation energy $D$ separates the ground state with spin magnetic quantum number $M_S=0$ from the degenerate excited spin states with $M_S=\pm 1$ (Fig.\,\ref{fig1}b)\@.

The exchange interaction between the Nc tip and various magnetic adsorbates on surfaces has successfully been probed by the sensitive response of the Nc spin excitation spectrum \cite{science_364_670,science_366_623,acsnano_16_16402, acsnano_18_13829}.
In addition, the spin state of Nc at the STM tip can be manipulated by bringing the Nc tip close to the surface \cite{natcommun_8_1974,prb_101_075414}.
The transition from $S=1$ to $S=1/2$ is reflected by the abrupt change of the spin excitation gap in spectroscopy of the differential conductance ($\text{d}I/\text{d}V$, $I$: current, $V$: sample voltage) to a zero-voltage resonance that is assigned to the Kondo effect \cite{natcommun_8_1974}.
A gradual closing of the spin excitation gap was likewise observed \cite{prb_101_075414}, albeit less frequently.

While Nc-decorated probes have been demonstrated to act as sensitive spin sensors when nearly in contact with surfaces, the impact of the atomic structure of the electrodes -- tip and surface -- on the Nc spin states has experimentally not been addressed so far.
Therefore, the dependence of the spin excitation spectrum is explored here on the specific atomic environment of Nc adsorbed on Cu(111), a single Cu atom and an artificially fabricated linear Cu$_3$ cluster.
In addition, variations of the spin excitation spectra for a single Nc decorating a blunt STM tip approaching the surface and an adsorbed atom (adatom) are compared.
The experimental observations are manifold.
Spin excitation energies depend on the atomic sharpness of the electrode Nc is attached to, with a maximum energy attained for the single-atom-terminated electrode.
Moreover, for the approach to Cu(111) the gradual decrease of the spin excitation energies and the eventual evolution of a zero-voltage resonance in $\text{d}I/\text{d}V$ spectroscopy at Nc--surface contact is observed.
In contrast, the experiments with the same Nc tip on the adatom lead to the sudden disappearance of the spin excitation gap for the benefit of a $\text{d}I/\text{d}V$ zero-voltage resonance.
These findings furthermore unveil the hitherto elusive mechanism for gradual and abrupt changes in Nc spin excitation spectra \cite{prb_101_075414}.

Stable imaging of single Nc molecules with STM was hampered by perpetual lateral translations of the molecules in the presence of the tip, which is indicative of a weakly corrugated adsorption potential for Nc on Cu(111)\@.
Only at surface defect sites, e.\,g., bulk-segregated impurities or lattice defects, Nc resides in a manner that enables stable STM imaging as shown in Fig.\,\ref{fig1}c.
The broad ring with a central depression is assigned to the top Cp moiety of an upright adsorbed Nc, in agreement with observations from Ag(110) \cite{science_364_670} and Cu(100) \cite{prb_93_195403,nl_17_1877,science_366_623,natcommun_8_1974,prb_101_075414,acsnano_16_16402}.
Such Nc molecules can readily be transferred to the tip by reducing the separation between the tip and the Nc center at negative sample voltage ($-1\,\text{mV}$, Supporting Information, Sec.\,S1)\@.
The successful transfer can be verified by imaging an atom adsorbed on the surface, which represents an elegant method to geometrically characterize the atomic-scale details of the tip \cite{jvstb_14_593}.
A Cu adatom appears with a ringlike structure in STM images acquired with the Nc-decorated tip (Fig.\,\ref{fig1}d)\@.
The asymmetric brightness of the ring and the off-center depression indicate a tilted axis of the Nc molecule terminating the tip, which is estimated as $\approx 6^\circ$ with respect to the surface normal (Supporting Information, Sec.\,S2)\@.

In addition to imaging, spectroscopy of $\text{d}I/\text{d}V$ with the Nc-terminated tip was performed on Cu(111) (top spectrum in Fig.\,\ref{fig1}e) and atop a Cu adatom (bottom)\@.
Both spectral data were recorded in the far tunneling range of tip--surface distances and exhibit a virtually identical gap with flanks symmetrically positioned around zero sample voltage, which are attributed to the spin excitation from $M_S=0$ to $M_S=\pm 1$, i.\,e., to the spin flip from the horizontal to the vertical spin orientation.
Matching the data with $f(V)=\sigma_0 \cdot\left[\Xi(\text{e}V-D,T)+\Xi(-\text{e}V-D,T)\right]+\sigma_{\text{el}}$, which results from the second-order scattering model \cite{njp_17_063016} and where $\Xi$ is the temperature-broadened step function \cite{pr_165_821} with $\sigma_0$, $D$, $T$, $\sigma_{\text{el}}$ serving as fit parameters, yields $D=3.7\pm 0.2\,\text{meV}$, which is in agreement with earlier results \cite{nl_17_1877}.
The values represent the arithmetic mean and standard deviation obtained from ten different Nc tips.

\begin{figure}
\centering
\includegraphics[width=\textwidth]{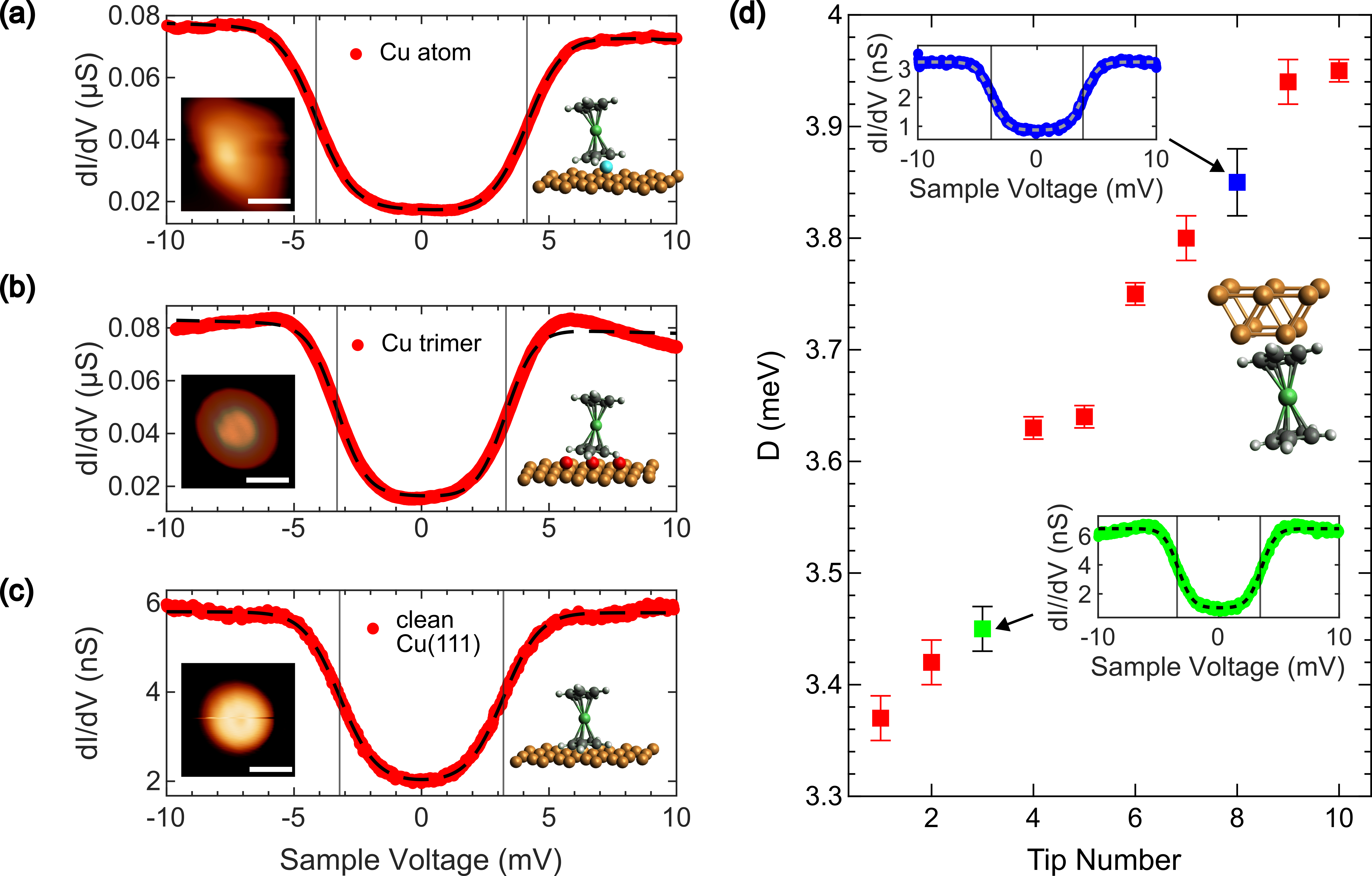}
\caption{Influence of the electrode structure on the Nc spin excitation energy.
(a) Spectrum of $\text{d}I/\text{d}V$ acquired with a pristine Cu tip atop the center of a single Nc molecule adsorbed on a single Cu atom on Cu(111) (feedback loop parameters: $10\,\text{mV}$, $500\,\text{pA}$)\@.
Insets: STM image of adsorbed Nc ($80\,\text{mV}$, $20\,\text{pA}$) and sketch of the suggested adsorption geometry.
(b) As (a), for Nc adsorbed on a linear Cu$_3$ cluster ($10\,\text{mV}$, $600\,\text{pA}$)\@.
(c) As (a), for Nc adsorbed on bare Cu(111) ($10\,\text{mV}$, $50\,\text{pA}$)\@.
The scale bars in all STM images indicate $1\,\text{nm}$.
(d) Collection of $D$ extracted from spin excitation spectra acquired in the far tunneling range ($10\,\text{mV}$, $30\,\text{pA}$--$100\,\text{pA}$) with $10$ Nc-decorated tips atop Cu(111)\@. 
The uncertainty margins mark variations of the second-order fit algorithm.
Insets: exemplary $\text{d}I/\text{d}V$ spectra associated with the marked data points and sketch of suggested Nc decoration of a blunt tip.
The vertical lines in all spectra mark $V=\pm D/\text{e}$, with $D$ resulting from fits (dashed lines) to the $\text{d}I/\text{d}V$ data within the second-order scattering model (Supporting Information, Sec.\,S7 and Fig.\,S6)\@.}
\label{fig2}
\end{figure}

The scattering of $D$ is assigned to the atomic-scale environment Nc is embedded in, which is an important finding of the studies presented here and which helps elucidate the evolution of the Nc spin excitation spectra upon contact to be discussed below.
To see the impact of the atomic electrode geometry on the spin excitation energy, Nc-decorated tips were mimicked by Nc molecules adsorbed on a single Cu atom on Cu(111), on an assembled linear Cu$_3$ trimer, and on Cu(111), which model atomically sharp, less sharp and blunt tip configurations, respectively.
Single Cu atoms were transferred from the tip to the surface \cite{prl_94_126102}, while Cu$_3$ chains were assembled by atom manipulation (Supporting Information, Sec.\,S3 and Fig.\,S1)\@.
In agreement with previous reports \cite{prl_91_206102,prl_92_056803}, Cu dimers and compact Cu trimers were unstable at the experimental temperature.

The associated spin excitation spectra are depicted in Fig.\,\ref{fig2}a--c and show that the maximum $D$ is obtained for Nc residing on the adatom ($D_{\text{max}}\approx 4.14\,\text{meV}$)\@.
The excitation energy decreases to $D\approx 3.31\,\text{meV}$ for Nc on linear Cu$_3$ and reaches its minimum of $D_{\text{min}}\approx 3.21\,\text{meV}$ on pristine Cu(111)\@.
It is noteworthy that spectroscopy of Nc on clean Cu(111) was feasible only in the far tunneling range due to the aforementioned junction instabilities, while molecules anchored at surface defects (Fig.\,\ref{fig1}c) did not indicate spin excitation signatures.
Therefore, the spin excitation energy of Nc in these different contact junctions can be varied by nearly $1\,\text{meV}$, which corresponds to a modification of $D$ by $\approx 30\,\%$\@.

The entire set of ten Nc tips that were prepared for the experiments gave rise to spin excitation spectra with energies ranging between $D_{\text{min}}$ and $D_{\text{max}}$ (Fig.\,\ref{fig2}d)\@.
In particular, based on these results the Nc tip used in the experiments discussed next is proposed to belong to the class of rather blunt tips because the extracted $D\approx 3.45\,\text{meV}$ is close to $D_{\text{min}}$ (bottom inset to Fig.\,\ref{fig2}d)\@.

\begin{figure}
\centering
\includegraphics[width=\textwidth]{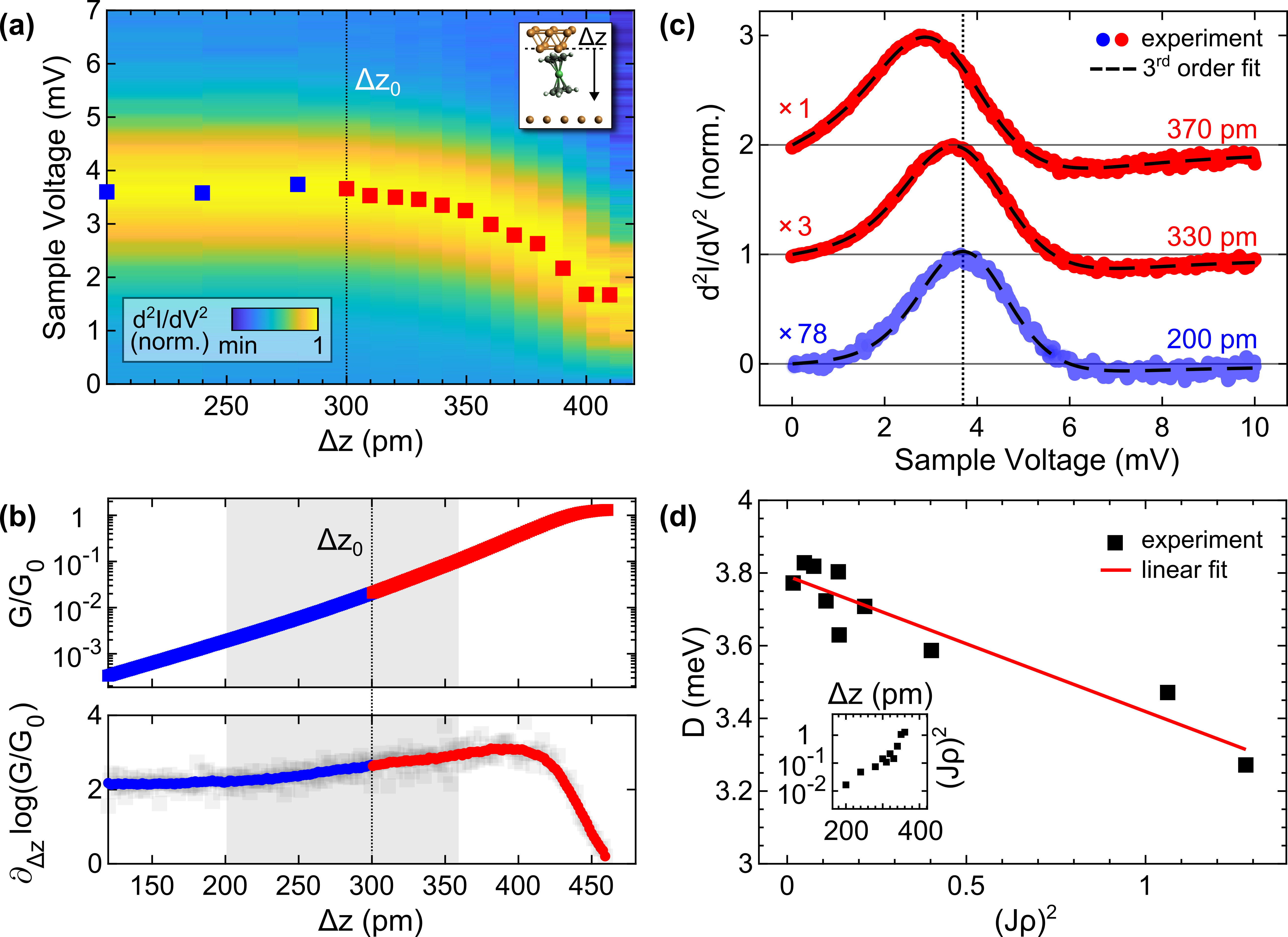}
\caption{Approach of an Nc tip to Cu(111)\@.
(a) Collection of $\text{d}^2I/\text{d}V^2$ spectra as a function of tip excursion $\Delta z$.
The spectra are normalized to their respective maxima. 
Symbols mark the $\text{d}^2I/\text{d}V^2$ peak positions. 
At $\Delta z_0\approx 300\,\text{pm}$ the shift of the spin excitation peaks toward $0\,\text{V}$ becomes detectable.
Feedback loop parameters defining tip excursion $\Delta z=0$: $10\,\text{mV}$, $20\,\text{pA}$\@.
(b) Top: variation of $G/\text{G}_0$ at $10\,\text{mV}$ as a function of tip approach (from low to high $\Delta z$) on a logarithmic scale.
Bottom: derivative of $\text{log}(G/\text{G}_0)$ with respect to $\Delta z$.
The shaded area marks the $\Delta z$ region where weak deviations of $G(\Delta z)$ from a uniform exponential behavior occur.
(c) Selected $\text{d}^2I/\text{d}V^2$ spectra (dots) at the indicated $\Delta z$ for illustrating the line shape analysis within the third-order dynamical scattering model (dashed lines)\@. 
The vertical line marks the position of the spin excitation peak in the far tunneling spectrum.
The spectra are vertically offset with the horizontal lines marking the respective zeros of $\text{d}^2I/\text{d}V^2$\@. 
(d) Evolution of $D$ with $(J\varrho)^2$ (squares) as extracted from the third-order fits for $200\,\text{pm}\leq\Delta z\leq 360\,\text{pm}$.
The solid line is a linear fit to the data.
Inset: variation of $(J\varrho)^2$ with $\Delta z$.}
\label{fig3}
\end{figure}

In a next step, the evolution of the spin excitation spectra atop Cu(111) and the Cu adatom was explored with decreasing separation between the Nc tip and the surface.
The collection of all $\text{d}^2I/\text{d}V^2$ spectra for $V\geq 0$ as a function of tip excursion $\Delta z$ on Cu(111) is depicted in Fig.\,\ref{fig3}a.
The spectra in the far tunneling range are dominated by a peak at $3.7\,\text{mV}$, which corresponds to the spin excitation of Nc (Fig.\,\ref{fig1}e)\@.
With increasing $\Delta z$, i.\,e., with decreasing tip--surface distance, the spin excitation signature shifts to lower sample voltages in a gradual manner, similar to observations on Cu(100) \cite{prb_101_075414}.
At the largest tip excursions used in the experiments the excitation peak appears at $1.7\,\text{mV}$\@.

To clearly see the collapse of the tunneling barrier at contact, traces of the junction conductance $G=I/V$ are useful \cite{jpcm_20_223001,pccp_12_1022}.
Figure \ref{fig3}b shows $G=G(\Delta z)$ in units of the conductance quantum $G_0=2\text{e}^2/\text{h}$ (e: elementary charge h: Planck constant) where the strongest deviation from a uniform exponential variation occurs starting from $\Delta z\gtrsim 420\,\text{pm}$.
The conductance trace then levels off and reaches a nearly constant value of $1.3\,\text{G}_0$, which is comparable to other single-molecule contacts \cite{jpcm_20_223001,pccp_12_1022,pssb_250_2437,csr_44_920,rmp_92_035001}.
This range of tip excursions is therefore associated with the Nc--Cu(111) contact.
Importantly, weak deviations from a uniform exponential $G(\Delta z)$ behavior are already observed for tip excursions indicated by the gray rectangle in Fig.\,\ref{fig3}b ($200\,\text{pm}\leq\Delta z\leq 360\,\text{pm}$)\@.
These changes are more clearly identified in the plot of the derivative $\partial\log(G/\text{G}_0)/\partial\Delta z$ in the bottom panel of Fig.\,\ref{fig3}b.
They are compatible with the shift of the spin excitation signature to lower bias voltage (Supporting Information, Sec.\,S4 and Fig.\,S2)\@.
Therefore, in this range of tip excursions, relaxations such as the deformation of the Nc backbone are negligible.

The evolution of the spin excitation can be understood from a careful analysis of its line shape in $\text{d}^2I/\text{d}V^2$ spectroscopy.
Figure \ref{fig3}c shows three exemplary spectra at different $\Delta z$.
Besides the peak shift to lower voltage with increasing $\Delta z$, the high-voltage tail of the peak exhibits a remarkable evolution.
In the far tunneling range of $\Delta z$ (bottom spectrum in Fig.\,\ref{fig3}c), the tail is virtually horizontal.
With increasing $\Delta z$, i.\,e., with increasing $G$, the tail evolves into a shallow and broad minimum.
The second-order tunneling model, which proved successful for describing the spin excitation spectra in the tunneling range (dashed lines in Fig.\,\ref{fig1}e), is therefore no longer applicable to the description of the spectral data at increased $G$ (Supporting Information, Sec.\,S5 and Fig.\,S3)\@.
Rather, higher-order physical effects have to be considered for explaining the actual variation of the $\text{d}^2I/\text{d}V^2$ signal with the voltage.

A possible rationale for the line shape evolution is the coupling of the Nc spin excitation to the substrate electron continuum via Kondo exchange interaction \cite{prb_97_075428}.
It is mediated by the hybridization of the magnetic impurity with the electron gas of the metal, which in the experiments is tuned by the proximity of the Nc-terminated tip to the surface.
Including the Kondo exchange interaction into the dynamical scattering model via the second-order Born approximation gives rise to logarithmic contributions in $\text{d}I/\text{d}V$ at $V=\pm D/\text{e}$ scaling with $J\varrho$ ($J$: Kondo exchange energy, $\varrho$: sample density of states at the Fermi energy)\@.
This extended third-order tunneling model \cite{njp_17_063016,prl_17_91,pr_154_633} matches the experimental data very well (Supporting Information, Sec.\,S5 and Fig.\,S4)\@.
Moreover, the experimentally extracted $D$ for $200\,\text{pm}\leq\Delta z\leq 360\,\text{pm}$ follows the predicted behavior, $D=D_0-c(J\varrho)^2$ ($D_0$: spin excitation energy without Kondo exchange coupling, $c$: constant factor containing the bandwidth of the relevant substrate electrons and the temperature) \cite{natnanotechnol_9_64,natcommun_6_8536}, as demonstrated in Fig.\,\ref{fig3}d.
Therefore, in agreement with observations and conclusions reported from Cu(100) \cite{prb_101_075414}, the gradual closing of the spin excitation gap is assigned to the increasing Kondo exchange interaction with the eventual Kondo screening of the $S=1$ impurity at contact.
In the case of Cu(100) the predicted variation of $D$ with $J\varrho$ is likewise applicable (Supporting Information, Sec.\,S6 and Fig.\,S5)\@.
Because the predicted behavior of $D$ with $J\varrho$ is well verified by the experimental data, a possible decrease of $D$ due to junction relaxations or charge transfer is surmised here to play a minor role.

\begin{figure}
\centering
\includegraphics[width=\textwidth]{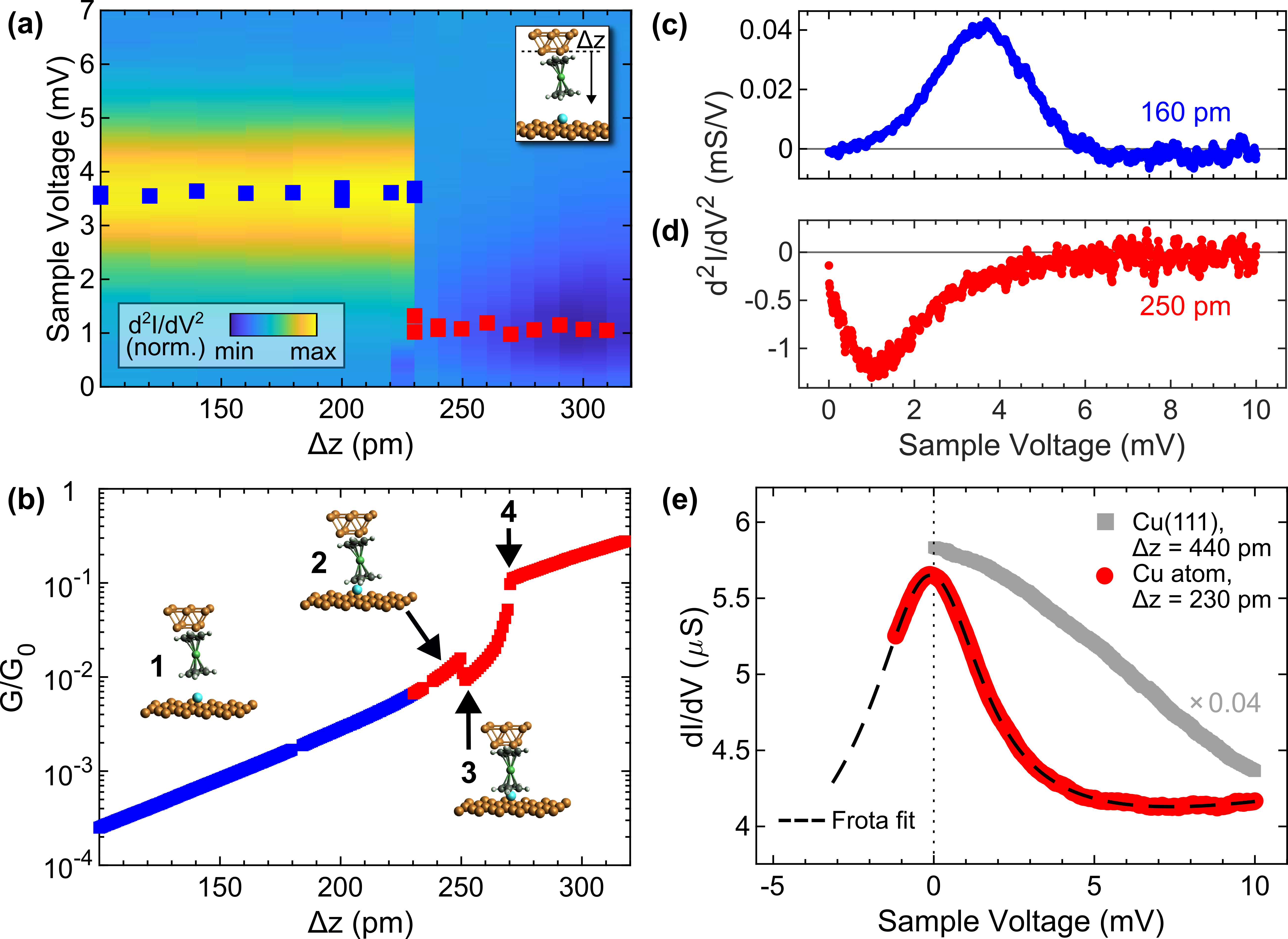}
\caption{Approach of the same Nc tip as in Fig.\,\ref{fig3} to a Cu adatom on Cu(111)\@.
(a) Collection of $\text{d}^2I/\text{d}V^2$ spectra as a function of tip excursion $\Delta z$.
Symbols mark peak and dip positions.
The spectra are normalized to the respective junction current at $10\,\text{mV}$. 
Feedback loop parameters defining $\Delta z=0$: $10\,\text{mV}$, $20\,\text{pA}$\@.
(b) Variation of the junction conductance at $10\,\text{mV}$ with $\Delta z$.
Insets: illustrations of suggested junction geometries in the tunneling ($\mathbf{1}$) and contact ($\mathbf{2}$--$\mathbf{4}$) ranges.
(c) Exemplary $\text{d}^2I/\text{d}V^2$ spectrum in the tunneling range ($\Delta z=160\,\text{pm}$)\@.
(d) As (c), for the contact range ($\Delta z=250\,\text{pm}$)\@.
(e) Integrated $\text{d}^2I/\text{d}V^2$ spectrum ($\Delta z=230\,\text{pm}$, dots) for an extended voltage range that was fit with a Frota function (dashed line)\@.
Gray symbols depict $\text{d}I/\text{d}V$ data obtained from the Nc--Cu(111) contact.}
\label{fig4}
\end{figure}

The same Nc tip that was used for the experiments on Cu(111) then likewise served as the probe for the Cu adatom (Fig.\,\ref{fig4})\@.
The Nc spin excitation occurs at a similar voltage ($\pm 3.7\,\text{mV}$) as on the pristine surface.
However, its evolution with increasing tip excursion is markedly different (Fig.\,\ref{fig4}a)\@.
Upon entering the contact range, which is signaled by the deviation of the junction conductance from its uniform exponential increase (\textbf{1} in Fig.\,\ref{fig4}b), the excitation peak disappears and a dip occurs instead in $\text{d}^2I/\text{d}V^2$ spectra.

A typical $\text{d}^2I/\text{d}V^2$ spectrum in the tunneling range is depicted in Fig.\,\ref{fig4}c.
It bears resemblance to the spectrum recorded atop Cu(111) (Fig.\,\ref{fig3}c)\@.
At contact, however, a dip occurs in the spectroscopic data (Fig.\,\ref{fig4}d), whose origin shall be clarified in the following.
To this end, the voltage range was extended to $-2\,\text{mV}$\@.
Below, junction instabilities occurred that hampered spectroscopy of the junction current.
A representative contact spectrum with the extended voltage range was numerically integrated giving rise to a peak centered at nearly $0\,\text{V}$ (Fig.\,\ref{fig4}e)\@.
Compared with integrated $\text{d}^2I/\text{d}V^2$ spectra of Cu(111) at comparable tip--surface distance (gray line in Fig.\,\ref{fig4}e), the line width of the peak for the adatom is considerably narrower.
The Frota line shape, $F(V)=a_2\cdot\Im\left[\text{i}\exp(\text{i}\varphi)\sqrt{\text{i}\Gamma/(\text{e}V-\varepsilon_0+\text{i}\Gamma)}\right]+a_1\cdot V+a_0$ ($\varphi$: asymmetry factor, $\Gamma$: resonance width, $\varepsilon_0$: resonance energy, $\text{i}^2=-1$) best matches the adatom data (dashed line in Fig.\,\ref{fig4}e). 
To account for thermal and voltage modulation broadening, the Frota peak was additionally convoluted with the associated functions \cite{prb_7_2336,jpcm_30_424001}.
Therefore, and because of previous findings on Cu(100) \cite{natcommun_8_1974,prb_101_075414}, it is reasonable to assume that the peak in $\text{d}I/\text{d}V$ at contact represents the Abrikosov-Suhl (AS) resonance, which signals the Kondo effect of an $S=1/2$ impurity \cite{prb_45_1096}.
In this case, the Kondo temperature results as \cite{prb_79_085106,natphys_7_203} $T_\text{K}=\Gamma/(1.455\,\text{k}_\text{B})=8\pm 1\,\text{K}$ ($\text{k}_\text{B}$: Boltzmann constant) and the resonance energy $\varepsilon_0$ essentially vanishes.
Based on the experimental results, the abrupt change in the spectral data is related to a transition of the Nc spin state from $S=1$ to $S=1/2$\@.
The former requires energy $D$ for a spin flip and, therefore, prevents the Kondo effect \cite{natphys_4_847}, while the latter exhibits degenerate electron spin levels \cite{natcommun_8_1974} where Kondo spin flip processes occur with vanishing energy cost.

The $G(\Delta z)$ trace (Fig.\,\ref{fig4}b) contains valuable hints to the mechanism underlying the spin transition. 
Unlike the behavior observed from approaching Cu(111), the $G(\Delta z)$ data for the adatom exhibits kinks.
The accelerated increase of $G(\Delta z)$ (left arrow in Fig.\,\ref{fig4}b) is likely related to the verge of the Nc--adatom contact, which is preferably formed with the Cp C--C bond (\textbf{2} in Fig.\,\ref{fig4}b) \cite{natcommun_8_1974,prb_101_075414}.
The subsequent conductance decrease (middle arrow in Fig.\,\ref{fig4}b) may be associated with the centering of the Cu adatom in the Cp ring of Nc, which is achieved by a straightening of the molecule in the junction (\textbf{3} in Fig.\,\ref{fig4}b)\@.
A similar scenario was previously put forward for a Cu adatom on Cu(100) mimicking a single-atom tip in simulations \cite{natcommun_8_1974}.
Further increase of $\Delta z$ is accompanied by a rapid rise of $G$ (\textbf{3} to \textbf{4} in Fig.\,\ref{fig4}b), which transitions into a uniform increase at \textbf{4}.
Presumably, at \textbf{4} an Nc-adatom contact configuration is reached that is not subject to strong relaxations with increasing $\Delta z$.
This $G$ region reliably signaled the spin transition, i.\,e., the presence of the zero-voltage resonance in $\text{d}I/\text{d}V$ spectra (Fig.\,\ref{fig4}e)\@.
Importantly, imaging the adatom and probing the spin excitation gap of the Nc tip atop the bare surface ensured the structural integrity of the junction.
These control experiments were intermediately performed during the entire approach experiment. 

The remaining question concerns the experimentally observed transition behavior of the spin excitation spectra, i.\,e., the gradual-versus-abrupt evolution of the excitation gap into a peak in $\text{d}I/\text{d}V$ spectroscopy.
The experiments reported here help clarify the situation because the contact to the adatom with a blunt Nc-terminated tip is accompanied by an abrupt transition of the Nc spin excitation gap into a $\text{d}I/\text{d}V$ zero-voltage peak that is similar to the observations reported from Cu(100) \cite{natcommun_8_1974,prb_101_075414} in absence of adatoms but, presumably, with atomically sharp Nc-terminated tips.
When the adatom is supposedly centered in the Cp ring upon straightening of Nc in the junction (\textbf{3} in Fig.\,\ref{fig4}b) the localized and protruding adatom orbitals can penetrate the Cp ring and hybridize with the Ni $d$-states.
The previously robust spin state $S=1$ is no longer protected by the C $p$-orbitals of the Cp moiety and transitions to $S=1/2$\@.
The hybridization of Nc orbitals and metal states were indeed demonstrated in density-functional simulations to impact the molecular magnetic moment and the spin excitation energy \cite{natcommun_11_1619,jpcl_13_11262}.
It is conjectured now that the straight Nc--atom junction may equivalently be formed at the tip apex.
In this suggested scenario the Cp group of Nc centers the Cu apex atom of the tip upon contact to the otherwise atomically flat substrate surface.
Previous calculations assuming a single-atom-terminated tip revealed this molecular relaxation at the tip in contact geometries, which entailed a decrease of the Ni magnetic moment from $1\,\mu_\text{B}$ to $0.5\,\mu_\text{B}$ ($\mu_\text{B}$: Bohr magneton) \cite{natcommun_8_1974}.
Consequently, the presence of the adatom in the single-Nc junction is required for the abrupt closing of the spectroscopic spin excitation gap, while its gradual disappearance is related to atomically flat electrodes.

In conclusion, the presented findings show that the spin excitation energy of a single magnetic molecule depends on its atomic environment.
Atomic-scale engineering of the electrode geometry the magnetic molecule is attached to, allows to tune the excitation energies by nearly $30\,\%$\@.
The hitherto elusive mechanism underlying the gradual and abrupt closure of the spin excitation energy gap has been related to, respectively, flat and atomically sharp electrodes attached to the molecule.
The results of this work are important for spintronics devices that rely on the stable spin structure, e.\,g., in data storage applications.
From a quantum-physics and quantum-chemistry perspective the work unveils that a spin state protected by a molecular matrix can be altered by atomic orbitals protruding into the matrix and hybridizing with the spin-carrying energy levels.

\section{Methods}
The experiments were performed with an STM operated in ultrahigh vacuum ($10^{-9}\,\text{Pa}$) and at low temperature ($5\,\text{K}$)\@.
Surfaces of Cu(111) were cleaned and prepared by Ar$^+$ ion bombardment and annealing.
The clean surface was exposed at a temperature between $5\,\text{K}$ and $20\,\text{K}$ to Nc molecules sublimated from a powder (purity: $\geq 98.5\%$) in a ceramics crucible held at room temperature. 
During Nc deposition the pressure rose to $5\cdot10^{-7}\,\text{Pa}$. 
A coverage of approximately $20$ Nc molecules per $80\,\text{nm}\times 80\,\text{nm}$ was achieved for an evaporation time of $5\,\text{min}$. 
A chemically edged W wire (purity: $99.95\,\%$, diameter: $50\,\mu\text{m}$) served as the tip material.
Field emission on and repeated indentations into the Cu surface presumably led to the coating of the tip apex with substrate material.
Single-atom transfer from the tip to the sample gave rise to particularly sharp and stable probes \cite{prl_94_126102,jpcm_20_223001,prl_102_086805,pccp_12_1022}.
Topographic data were acquired in the constant-current mode with the bias voltage applied to the sample and were further processed with WSxM \cite{rsi_78_013705}.
Spectroscopy of $\text{d}I/\text{d}V$ and $\text{d}^2I/\text{d}V^2$ proceeded via the sinusoidal modulation ($250\,\mu\text{V}$, $726\,\text{Hz}$ for $\text{d}I/\text{d}V$ and $363\,\text{Hz}$ for $\text{d}^2I/\text{d}V^2$) of the dc sample voltage and measuring, respectively, the first and second harmonic of the ac current response of the tunneling junction with a lock-in amplifier.

\section{Associated Content}

\subsection{Supporting Information} 
Supporting Information is available free of charge on the ACS Publication website at DOI: [hyperlink DOI] 

Tip termination, tilt angle of Nc-decorated tip, fabrication of molecule-cluster assemblies, junction relaxations, dynamical scattering models, Kondo exchange renormalization, third-order analysis applied to Nc tips in the far tunneling range (PDF)

\subsection{Data availability statement} The authors declare that relevant data supporting the findings of this study are available on request. 

\subsection{Author contributions}
M.K. and N.N. carried out the experiments. 
M.K. analyzed the experimental data. 
M.K and J.K. wrote the manuscript with comments of all authors. 
L.L. and J.K. conceived the experiments.

\begin{acknowledgement}
Funding by the Deutsche Forschungsgemeinschaft (grant no.\ KR 2912/21-1) and the Agence Nationale de la Recherche (grant no.\ ANR-23-CE09-0036-01) as well as the German Federal Ministry of Education and Research within the ''Forschungslabore Mikroelektronik Deutschland (ForLab)'' initiative is acknowledged. 
We are indebted to Alexander Weismann and Richard Berndt for providing raw spectroscopy data obtained from Cu(100)\@.
Discussions with Marie-Laure Bocquet, Roberto Robles and Lorenz Meyer are appreciated.
\end{acknowledgement}

\providecommand{\latin}[1]{#1}
\makeatletter
\providecommand{\doi}
  {\begingroup\let\do\@makeother\dospecials
  \catcode`\{=1 \catcode`\}=2 \doi@aux}
\providecommand{\doi@aux}[1]{\endgroup\texttt{#1}}
\makeatother
\providecommand*\mcitethebibliography{\thebibliography}
\csname @ifundefined\endcsname{endmcitethebibliography}
  {\let\endmcitethebibliography\endthebibliography}{}

\end{document}


\maketitle

\section{S1 Tip termination} 

The attachment of a single Nc molecule to the STM tip apex followed a previously reported procedure \cite{nl_17_1877}.
The tip was positioned above the center of an individual Nc molecule anchored at a Cu(111) surface defect.
The feedback loop was then disabled ($-40\,\text{mV}$, $20\,\text{pA}$) and the tip approached by $350\,\text{pm}$ at $-1\,\text{mV}$\@.
The successful transfer of Nc from the surface to the tip was verified by IETS of the Nc spin excitation on clean Cu(111)\@.
Additionally, the Nc-decorated tip apex was imaged by scanning across a single Cu atom that had been deposited prior to tip termination.
A similar approach to the spatial characterization of the tip apex was applied before \cite{jvstb_14_593,natnanotechnol_6_23,nl_19_7845}.

\section{S2 Tilt angle of Nc-decorated tip} 

As shown in the main text, an Nc-terminated tip gives rise to a ringlike pattern in the STM image of a single adsorbed Cu atom with nonunifom contrast (Fig.\,1d)\@.
The maximum apparent height difference $\Delta h$ observed in the ring is assigned to the tilted adsorption of Nc at the tip.
The tilt angle $\alpha$ subtending the surface normal can be estimated from $\alpha=\tan^{-1}(\Delta h/\Delta x)$, where $\Delta x$ is the lateral distance spanning the minimum and maximum apparent height of the ring.
The tips used in the present experiments exhibited $\alpha\approx 6^{\circ}$, which is in agreement with previous findings \cite{science_364_670,science_366_623}.

\section{S3 Fabrication of molecule-cluster assemblies} 

\begin{figure}
\includegraphics[width=0.4\columnwidth]{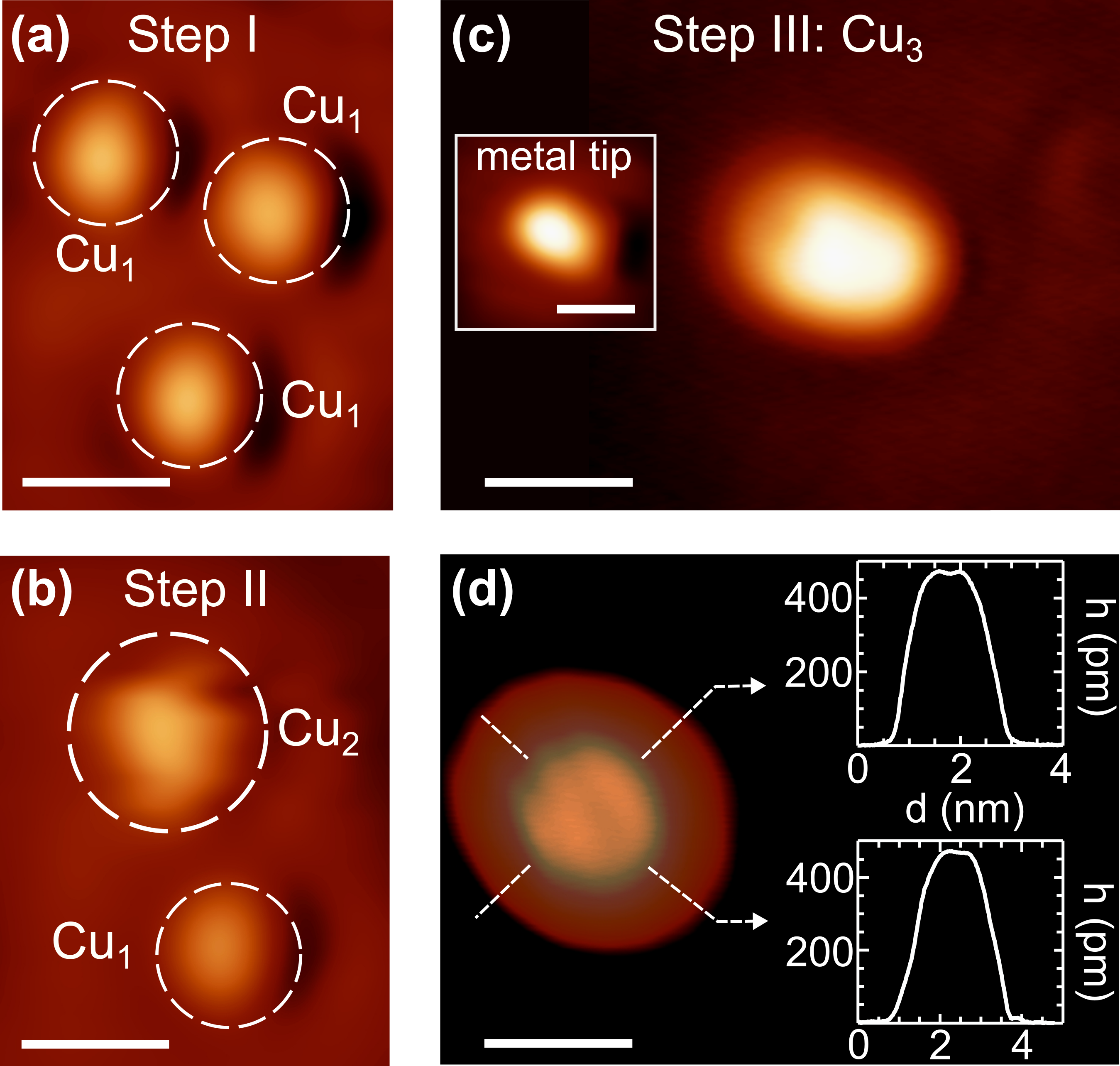}
\caption{Fabrication of Nc-cluster assemblies.
(a) STM image of three individual Cu atoms on Cu(111) acquired with a metal tip ($80\,\text{mV}$, $20\,\text{pA}$)\@.
(b) As (a) after Cu$_2$ assembly.
(c) STM image of a linear Cu$_3$ cluster on Cu(111) acquired with an Nc tip ($-200\,\text{mV}$, $20\,\text{pA}$)\@.
Inset: metal-tip STM image of the Cu$_3$ cluster in (c)  ($80\,\text{mV}$, $20\,\text{pA}$)\@.
(d) STM image of Nc-Cu$_3$ recorded with a metal tip ($80\,\text{mV}$, $20\,\text{pA}$)\@.
Insets: cross-sectional profiles acquired atop Nc-Cu$_3$ along the dashed lines.
The scale bars in all STM images indicate $1\,\text{nm}$.}
\label{figS1}
\end{figure}

In the first step of the manipulation experiments, a single Cu atom is transferred from the tip to the surface (Fig.\,\ref{figS1}a) according to a previously reported method \cite{prl_94_126102}. 
In the second step, one of the three deposited Cu atoms is dragged to another Cu atom to form a Cu$_2$ cluster (Fig.\,\ref{figS1}b)\@.
To this end, the metal tip is approached to the single Cu atom until the junction conductances reaches $\approx\text{G}_0/8$ at $10\,\text{mV}$ \cite{science_319_1066}.
The resulting Cu dimer is unstable at the temperature of the experiments and can be easily distorted by the imaging process (Fig.\,\ref{figS1}b), which agrees with previous findings \cite{prl_91_206102,prl_92_056803}.
Subsequently, the remaining Cu atom is dragged close to the Cu$_2$ cluster and a Cu$_3$ chain forms (Fig.\,\ref{figS1}c)\@.
The linear atomic assembly is preferred to the compact arrangement, which likewise is in accordance with earlier results \cite{prl_91_206102,prl_92_056803}.
The chain is oriented along a compact Cu(111) lattice direction and individual Cu atoms are adsorbed on face-centered cubic sites \cite{prl_91_206102,prl_92_056803}.

After the successful cluster assembly, the tip is decorated with a single Nc molecule, as described in Sec.\,S1\@.
The subsequent attachment of Nc to the pristine Cu(111) surface, the Cu adatom, and the Cu$_3$ cluster proceeds via the approach of the Nc-terminated tip.
For Cu(111), the feedback loop was deactivated at $-40\,\text{mV}$, $20\,\text{pA}$, for the Cu adatom and the Cu$_3$ cluster at $10\,\text{mV}$, $20\,\text{pA}$ prior to approaching the tip by $450\,\text{pm}$ at $50\,\text{mV}$\ (surface), $200\,\text{pm}$ (adatom) and $250\,\text{pm}$ (Cu$_3$) at $10\,\text{mV}$\@. 
These data were found to scatter, which hints at the dependence of the assembly on the actual tip\@.

The Nc-adatom compound can easily be distorted by imaging (inset to Fig.\,2a of the article), in agreement with observations from Nc-adatom compounds on Cu(100) \cite{prb_101_075414}. 
One stable junction was achieved and reproducible dI/dV spectra of the spin excitation gap could be acquired, that is, two consecutive measurements, each performed after a control image yielded the same $D=4.14\,\text{meV}$.

The Nc-Cu$_3$ assembly is stable (Fig.\,\ref{figS1}d)\@.
Cross-sectional profiles (insets to Fig.\,\ref{figS1}d) show that Nc is slightly tilted along the long axis of the Cu$_3$ chain. 
The stability allowed spectroscopic data acquisition from the tunneling to the contact range of tip separations. 
Spectroscopy was performed for one representative Nc-Cu$_3$. 
At a given tip--surface distance in the tunneling range, a reproducible $D=3.31\,\text{meV}$ resulted from three of such measurements. 
The associated standard deviation of only $0.01\,\text{meV}$ demonstrates the low scattering of the spectra.

\section{S4 Junction relaxations}

Two observations suggest that relaxation effects of the junction geometry play a minor role in the changes of the spin excitation energy above Cu(111)\@.
\begin{figure}
\centering
\includegraphics[width=0.5\columnwidth]{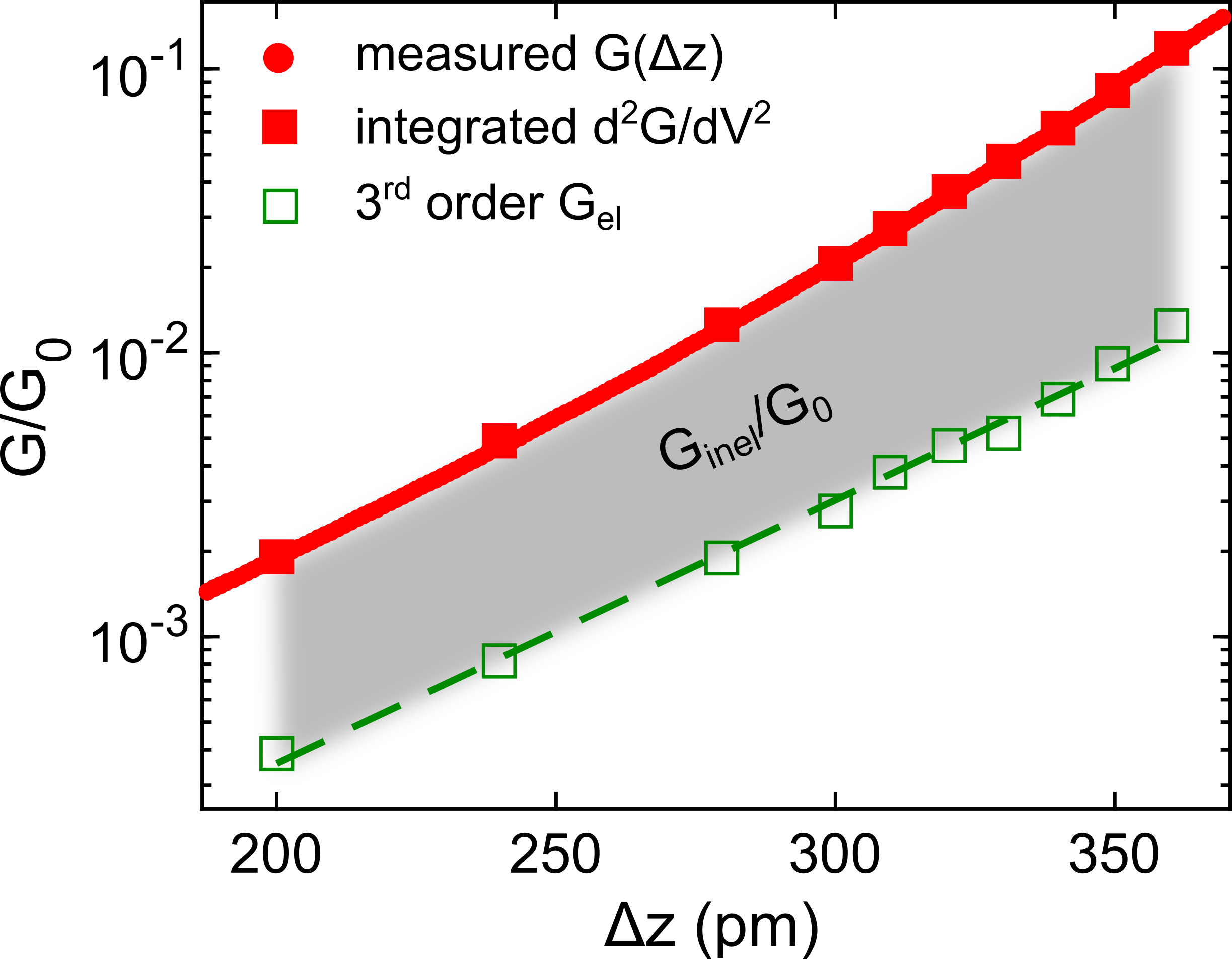}
\caption{Variation of junction conductance $G$ with Nc tip excursion $\Delta z$ on Cu(111) at $10\,\text{mV}$\@.
Measured $G(\Delta z)$ appears as dots (top graph), while integrated $\text{d}^2I/\text{d}V^2$ data (divided by $10\,\text{mV}$) are depicted as solid squares (integration interval: $0\leq V\leq 10\,\text{mV}$)\@.
Elastic contribution $G_{\text{el}}$ (open squares) to $G$ results from matching $\text{d}^2I/\text{d}V^2$ data within the third-order model.
The green dashed line depicts an exponential fit to $G_{\text{el}}$ data.
The shaded area illustrates the inelastic contribution $G_{\text{inel}}$ to $G$\@.}
\label{figS2}
\end{figure}
First, the variation of $D=D(J\varrho)$ matches previous calculations (Fig.\,4d) where mechanical deformations of the magnetic impurity are absent \cite{wiley_2008,natnanotechnol_9_64, natcommun_6_8536}.
Second, deviations of $G=G(\Delta z)$ (Fig.\,4b) from a uniform exponential variation, which are often assigned to junction relaxations, can readily be explained by the shift of the spin excitation onset to lower voltages in $\text{d}I/\text{d}V$ spectra.
To see this, Fig.\,\ref{figS2} compares the measured $G(\Delta z)$ (dots) with numerically integrated $\text{d}^2G/\text{d}V^2$ spectroscopic data (solid squares)\@.
Both graphs match very well.
In order to rule out elastic effects from contributing to the shifted spectroscopic signature, the elastic component $G_{\text{el}}$ of $G$ resulting from $\sigma_{\text{el}}$ in the third-order scattering model is added to Fig.\,\ref{figS2} (open squares)\@.
Obviously, $G_{\text{el}}$ virtually follows a uniform exponential variation, that is, elastic effects do not contribute to the buckled evolution of $G(\Delta z)$\@.

\section{S5 Dynamical scattering models}

\paragraph{Second order.} 

\begin{figure}
\centering
\includegraphics[width=\columnwidth]{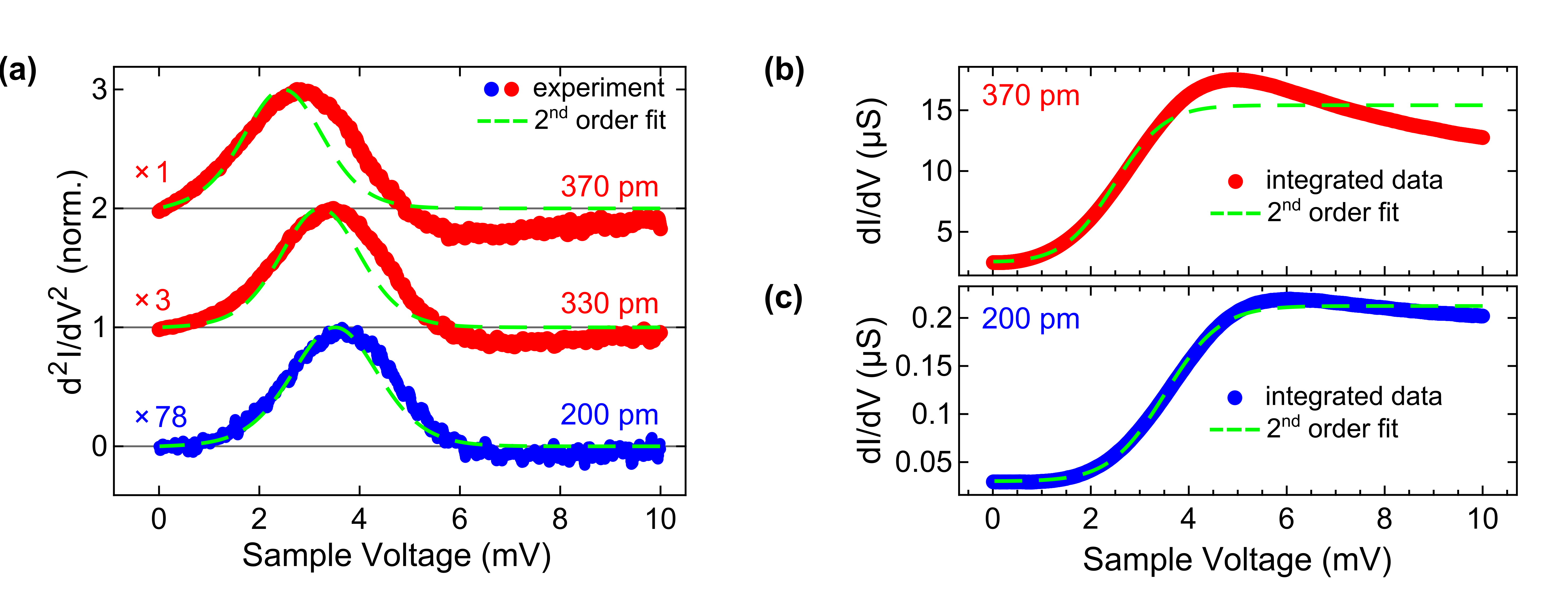}
\caption{Failure of the second-order model.
(a) Spectra of $\text{d}^2I/\text{d}V^2$ (dots) acquired with an Nc tip atop Cu(111) at the indicated tip excursions and normalized to the respective maxima (reference feedback loop parameters: $10\,\text{mV}$, $20\,\text{pA}$)\@.
Fits to the data are depicted as dashed lines.
The spectra and fits are vertically offset with the horizontal line marking $\text{d}^2I/\text{d}V^2=0$ for each spectrum.
(b) Numerically integrated data of (a) at $\Delta z=370\,\text{pm}$.
(c) Numerically integrated data of (a) at $\Delta z=200\,\text{pm}$.}
\label{figS3}
\end{figure}

A first approach to model the Nc spin excitation spectrum considers the isolated exchange interaction between the spin $\mathbf{s}$ of the tunneling electron and the molecular spin $\mathbf{S}$. 
The scattering potential is proportional to $\mathbf{s}\cdot\mathbf{S}$\@. 
Writing the differential conductance as $\text{d}I/\text{d}V\equiv\sigma=\sigma_{\text{el}}+\sigma_{\text{inel}}$ with elastic (el) as well as inelastic (inel) contributions and applying Fermi's Golden Rule to the aforementioned scattering process leads to in second order \cite{njp_17_063016}
\begin{equation}
\sigma_{\text{inel}}\equiv\sigma_{\text{inel}}^{(2)}=\sigma_0\sum\limits_{if}p_i\left|\mathbf{M}_{if}\right|^2\Xi(\text{e}V-\varepsilon_{if},T)
\label{SI_2ndOrder}
\end{equation}
($\ket{i}$,$\ket{f}$: all possible initial and final spin states of Nc, $p_i$: initial-state occupation, $\mathbf{M}_{if}$: transition matrix element, $\varepsilon_{if}$: energy difference of specific initial and final spin states)\@. 
The temperature-broadened step function $\Xi$ reads \cite{pr_165_821}
\begin{equation}
\Xi(\text{e}V-\varepsilon_{if},T)=\frac{\exp(v)\cdot[\exp(v)-v-1]}{[\exp(v)-1]^2}
\end{equation}
with $v=(\text{e}V-\varepsilon_{if})/(\text{k}_{\text{B}}T)$ and $\text{k}_{\text{B}}$ the Boltzmann constant.
The modulation broadening \cite{prb_7_2336,pr_165_821} of $\text{d}I/\text{d}V$ data is negligible compared with the temperature broadening for the modulation voltages used in the experiments. 
The matching of experimental $\text{d}^2I/\text{d}V^2$ data (Fig.\,\ref{figS3}a) proceeds via the numerical derivative $\text{d}\sigma/\text{d}V$ within the second-order scattering model.

For sufficiently small currents, nonequilibrium population of spin states is suppressed and $p_i$ solely reflects the thermal occupation of the spin states, which at the experimental temperature $T=5\,\text{K}$ is the ground state $\ket{M_{S}=0}$ since $\text{k}_{\text{B}}T\ll D$\@. 
The matrix elements $\mathbf{M}_{if}$ for exchange scattering between the spin states of the scattering electron ($\varphi_{i'},\varphi_{f'}$) and Nc ($\psi_{i},\psi_{f}$) are defined as
\begin{equation}
\mathbf{M}_{if}=\sum\limits_{i',f'}\braket{\varphi_{f'},\psi_f|\hat{\mathbf{s}}\cdot\hat{\mathbf{S}}|\psi_i,\varphi_{i'}}
\label{Mif_def}
\end{equation}
with $\hat{\mathbf{s}}$ and $\hat{\mathbf{S}}$ the associated spin operators.
For the transition $\ket{i}\equiv\ket{M_{S}=0}\rightarrow\ket{f}\equiv\ket{M_{S}=\pm 1}$, the nonzero squared matrix elements are calculated as $|\mathbf{M}_{0,+1}|^2=1/2=|\mathbf{M}_{0,-1}|^2$ with $\varepsilon_{0,+1}=D=\varepsilon_{0,-1}$ giving rise to $f(V)$ introduced in the main text ($\hbar = 1$). 
The temperature $T$ and the spin excitation energy $D$ served as fit parameters.
The resulting temperature reliably reproduced the experimental value of $T=5\,\text{K}$\@.

In the far tunneling range of tip--surface distances, the second-order model is successful (Fig.\,1e in the main text), while it fails in describing the data acquired at small tip--surface separations. 
This failure is illustrated in Fig.\,\ref{figS3}a for fits to $\text{d}^2I/\text{d}V^2$ data at different tip--surface distances and in Fig.\,\ref{figS3}b,c for the $\text{d}I/\text{d}V$ signal.
The main deviation occurs at the high-voltage tails.
The shallow indentation following the peak in $\text{d}^2I/\text{d}V^2$ (Fig.\,\ref{figS3}a), which is associated with the cusps in $\text{d}I/\text{d}V$ (Fig.\,\ref{figS3}b,c) is characteristic for third-order tunneling processes explained in the next paragraph.

\paragraph{Third order.}

In the third-order scattering model, the Kondo exchange interaction $-J\varrho \, \tilde{\mathbf{s}}\cdot\mathbf{S}$ is introduced where the spin of substrate electrons ($\tilde{\mathbf{s}}$) enters into the scattering potential ($J$: Kondo magnetic exchange energy, $\varrho$: density of electron states at the Fermi energy ($E_{\text{F}}$)) \cite{prl_17_91,pr_154_633,njp_17_063016}.
Here, processes are included where, e.\,g., substrate electrons scatter the Nc spin into some virtual intermediate state before the tunneling electron transfers this state to the final state. 
Considering scattering processes up to third-order is associated with the following transition matrix element that couples initial ($i$) and final ($f$) spin states of Nc via intermediate ($m$) spin states \cite{prl_17_91,pr_154_633}
\begin{equation}
\left\vert\mathbf{M}_{if}^{(3)}\right\vert^2=\left\vert\mathbf{M}_{if}\right\vert^2+J\rho\sum\limits_{m}\int\frac{2\Re\left(\mathbf{M}_{mi}\mathbf{M}_{fm}\mathbf{M}_{if}\right)}{\tilde{\varepsilon}_{im'}+\varepsilon_{im}}\,\text{d}\tilde{\varepsilon}_{m'}
\label{Mif_total}
\end{equation} 
The sum in eq \ref{Mif_total} reflects discrete Nc spin states, while the integral considers the continuum of nearly-free substrate electron states with energy $\tilde{\varepsilon}_{m'}$.
Evaluating the transition matrix elements in the numerator of eq \ref{Mif_total} by using eq \ref{Mif_def} yields $\Re\left(\mathbf{M}_{mi}\mathbf{M}_{fm}\mathbf{M}_{if}\right)=-1/4$ as the only nonzero contribution for the transitions $\ket{i}\equiv\ket{M_S=0}\rightarrow\ket{m}\equiv\ket{M_S=1}\rightarrow\ket{f}\equiv\ket{M_S=1}$ and $\ket{i}\equiv\ket{M_S=0}\rightarrow\ket{m}\equiv\ket{M_S=-1}\rightarrow\ket{f}\equiv\ket{M_S=-1}$\@.

In analogy to eq \ref{SI_2ndOrder}, the inelastic differential conductance in the third-order scattering model is expressed as
\begin{equation}
\sigma_{\text{inel}}=\sigma_0\sum\limits_{if}p_i\left\vert\mathbf{M}_{if}^{(3)}\right\vert^2\Xi(\text{e}V-\varepsilon_{if},T) = \sigma_{\text{inel}}^{(2)}+\sigma_{\text{inel}}^{(3)}
\label{eq:third_order}
\end{equation}
\begin{figure}[t]
\centering
\includegraphics[height = 0.55\textheight]{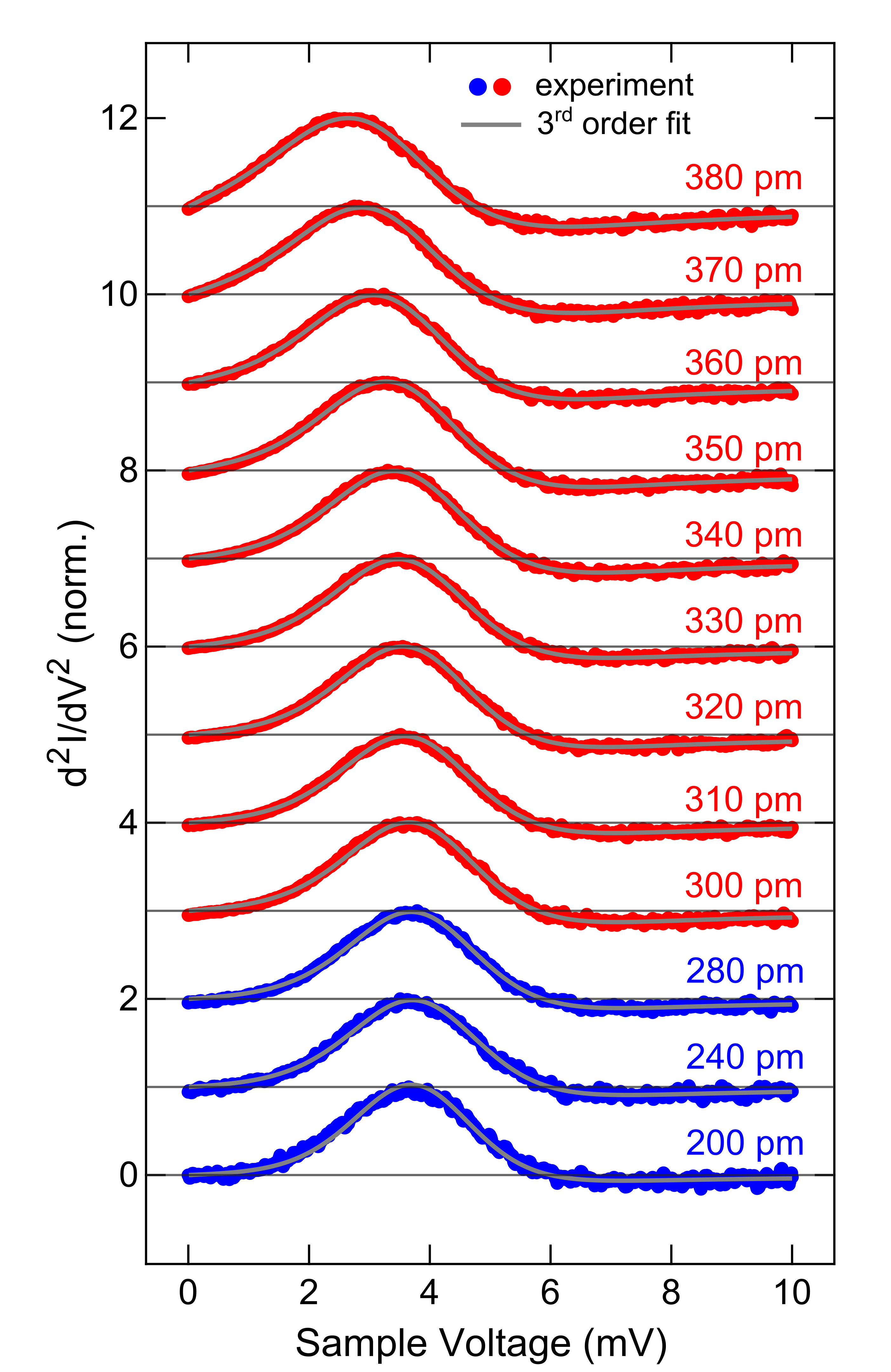}
\caption{Successful description of spectroscopic data within the third-order model.
Spectra of $\text{d}^2I/\text{d}V^2$ (dots) acquired with an Nc tip atop Cu(111) and normalized by the respective maximum at the indicated tip excursions $\Delta z$. Increasing $\Delta z$ corresponds to decreasing tip--sample distance.
The spectra are vertically offset with the horizontal lines marking $\text{d}^2I/\text{d}V^2=0$ for each tip excursion. Feedback loop parameters defining $\Delta z=0$: $10\,\text{mV}$, $20\,\text{pA}$\@.}
\label{figS4}
\end{figure}
Taking further into account the Fermi-Dirac occupation statistics $F(\tilde{\varepsilon})=1/\{1+\exp[(\tilde{\varepsilon}-E_{\text{F}})/(\text{k}_{\text{B}}T)]\}$ for initial, intermediate and final states, $\sigma_{\text{inel}}^{(3)}$ can be recast to give
\begin{equation}
\sigma_{\text{inel}}^{(3)}=-1\cdot\sigma_0\cdot J\rho\cdot\left[g(\text{e}V-D)+g(\text{e}V+D)\right]\cdot\left[\Xi(\text{e}V-D,T)+\Xi(-\text{e}V-D,T)\right] 
\label{eq:sigma3}
\end{equation} 
with  
\begin{equation}
g(\tilde{\varepsilon})=\int\limits_{-\infty}^{\infty}\int\limits_{-\hbar\omega_0}^{\hbar\omega_0}\frac{1-F(\tilde{\varepsilon}_{m'},T)}{\tilde{\varepsilon}_{m'}-\tilde{\varepsilon}_{i}+\text{i}\Gamma_0}F'(\tilde{\varepsilon}_{i}-\tilde{\varepsilon},T)\,\text{d}\tilde{\varepsilon}_{m'}\,\text{d}\tilde{\varepsilon}_i
\label{ln_t2}
\end{equation} 
($\text{i}^2=-1$, $\Gamma_0$: electron lifetime broadening) where $F'$ denotes the derivative of the Fermi-Dirac function with respect to $\tilde{\varepsilon}$ \cite{prl_17_91,pss_92_83,natcommun_6_8536}.
The inner integration is restricted to $-\hbar\omega_0\leq\tilde{\varepsilon}_{m'}-E_{\text{F}}\leq\hbar\omega_0$, which reflects the bandwidth of relevant substrate electron states \cite{prl_17_91,pr_154_633}. 

The fit of the resulting $\sigma=\sigma_{\text{el}}+\sigma_{\text{inel}}^{(2)}+\sigma_{\text{inel}}^{(3)}$ to the experimental $\text{d}I/\text{d}V$ data \cite{njp_17_063016} enables the extraction of $J\varrho$ and $D$, which served as fit parameters.
To this end, experimental $\text{d}^2I/\text{d}V^2$ data were corrected for offsets and then numerically integrated and symmetrized with respect to zero sample voltage.
The parameters $\Gamma_0$ and $\hbar\omega_0$ turned out to be dispensable for the overall quality of the fit and were therefore set to $\Gamma_0=5\,\mu\text{eV}$ and $\hbar\omega_0=20\,\text{meV}$ \cite{njp_17_063016, natcommun_6_8536}.
The fits reproduce the spectroscopic data for all measured tip--surface distances (Fig.\,\ref{figS4})\@.

\section{S6 Kondo exchange renormalization}

The measured spin excitation spectrum of Nc on Cu(111) reflects spin excitation energies $D$ that are renormalized due to the interaction of the Nc spin with the substrate electron continuum.
This Kondo exchange interaction was previously shown \cite{wiley_2008,natnanotechnol_9_64, natcommun_6_8536} to change the bare Nc spin energy levels according to $\varepsilon_{\alpha}(J\varrho)=\varepsilon_{\alpha}(0)+\delta\varepsilon_{\alpha}(J\varrho)$ with \cite{natcommun_6_8536}:
\begin{equation}
\delta\varepsilon_{\alpha}(J\varrho)\approx(J\varrho)^2\sum\limits_{n\neq\alpha}\sum\limits_{n',\alpha'}\frac{\left\vert\braket{\varphi_{n'},\psi_n| \hat{\tilde{\mathbf{s}}}\cdot\hat{\mathbf{S}}|\psi_{\alpha},\varphi_{\alpha'}}\right\vert^2}{\varepsilon_{\alpha}-\varepsilon_{n}+\tilde{\varepsilon}_{\alpha'}-\tilde{\varepsilon}_{n'}}
\label{E_i_Ren}
\end{equation}
From eq \ref{E_i_Ren}, it can be inferred that the Nc spin states $\ket{\alpha}\equiv\ket{M_S=\pm 1,0}$ move more closely in energy.
Consequently, the experimentally measured $D=\varepsilon_{0,\pm 1}$ is lowered.
Moreover, eq \ref{E_i_Ren} predicts the approximate behavior $D(J\varrho)\approx D_0-c(J\varrho)^2$ with $D_0=D(0)$ the spin excitation energy in the absence of Kondo exchange coupling and a constant $c$.
Since $D$ and $J\varrho$ are extracted from the matching procedure, this prediction can be tested.\\

\begin{figure}
\centering
\includegraphics[width=0.9\columnwidth]{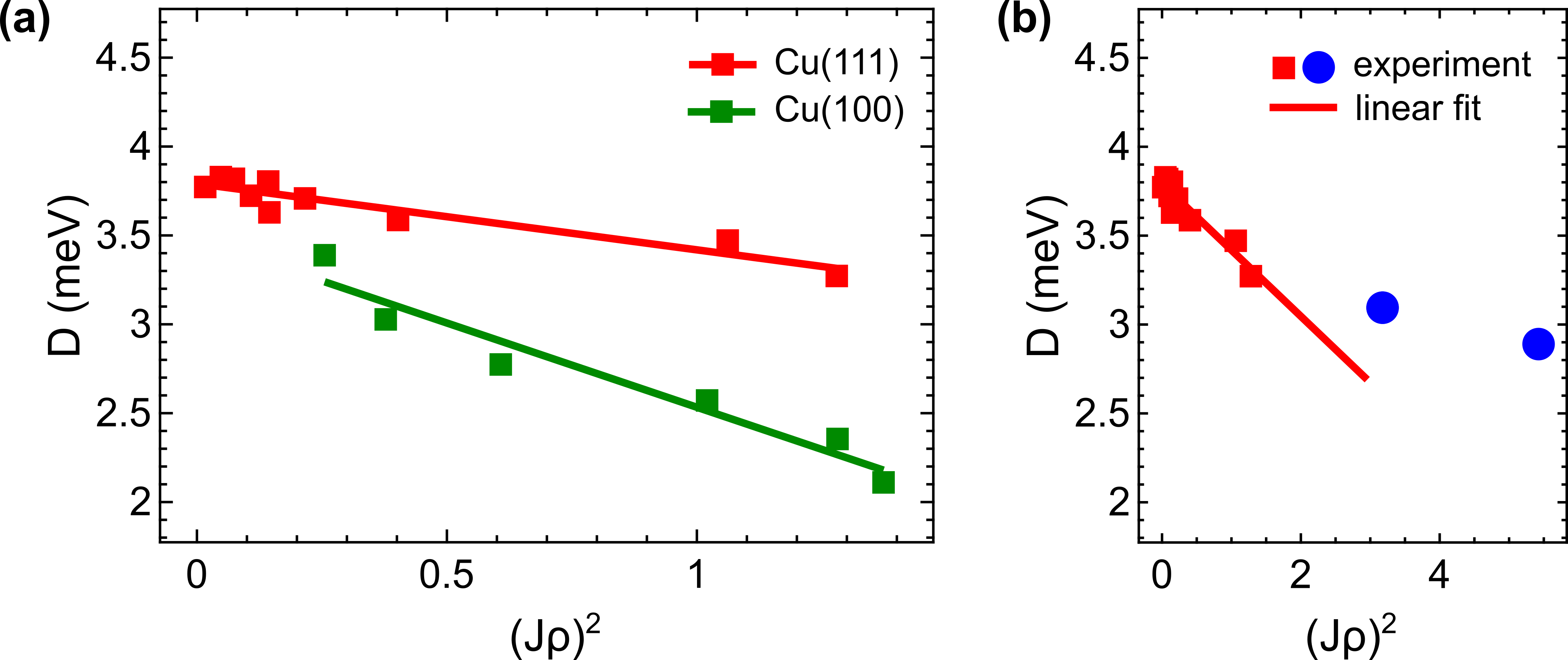}
\caption{Variation of $D$ with $(J\varrho)^2$ for an Nc tip on Cu(111) (this work) and Cu(100) \cite{prb_101_075414}.
(a) Green and red squares depict $D$ extracted from matching spectroscopic data within the third-order model while solid lines are linear fits to the data.
(b) Plot of Cu(111) data for an extended range of $J\varrho$ revealing the deviation (dots) from the predicted behavior $D(J\rho)$\@. The solid line is an extrapolated linear fit to the data presented in (a) (red squares).
The additional data were extracted from spectroscopic data acquired at tip excursions $370\,\text{pm}$ and $380\,\text{pm}$.}
\label{figS5}
\end{figure}

Figure \ref{figS5} shows a plot of $D$ as a function of $(J\varrho)^2$\@.
For sufficiently low $J\varrho$, $D$ indeed exhibits the predicted linear variation with $(J\varrho)^2$ (Fig.\,\ref{figS5}a)\@.
By courtesy of the Kiel group, experimental spin excitation spectroscopic data were provided for an Nc tip on Cu(100) \cite{prb_101_075414}.
Describing these data within the third-order scattering model likewise gives rise to a linear evolution of $D$ with $(J\varrho)^2$, albeit with a different slope.
While for Cu(111), $c=0.37 \,\text{meV}$ was obtained, the slope for the Cu(100) data is $c=1.00 \,\text{meV}$\@.
A clear-cut rationale for this observation is missing to date.
However, the slope $c$ is determined by the bandwidth $\hbar\omega_0$ of the substrate electrons Nc is coupled to \cite{natnanotechnol_9_64,natcommun_6_8536}.
Therefore, the different electronic structure of Cu(111) and Cu(100) may contribute to the deviations in $c$.
The different values for $D_0$, i.\,e., $D_0=3.8\,\text{meV}$ for Cu(111) and $D_0=3.5\,\text{meV}$ for Cu(100), are likely associated with the different tip shapes used in the separate experiments (Fig.\,2 of the main text)\@.

Figure \ref{figS5}b shows that for larger values of $J\varrho$ deviations from the predicted behavior of $D(J\varrho)$ occur, which reflect the limitations of the perturbation theory underlying the derivation of $\delta\varepsilon_\alpha$ (eq \ref{E_i_Ren}).
Indeed, the two additional data points in Fig.\,\ref{figS5}b were obtained for tip excursions $\Delta z=370\,\text{pm}$ and $\Delta z=380\,\text{pm}$, which are close to the collapse of the tunneling barrier (Fig.\,3b of the main text) and entailed by an increased interaction between the Nc tip and the surface.

\section{S7 Third-order analysis applied to Nc tips in the far tunneling range}

\begin{figure}
\centering
\includegraphics[width=\columnwidth]{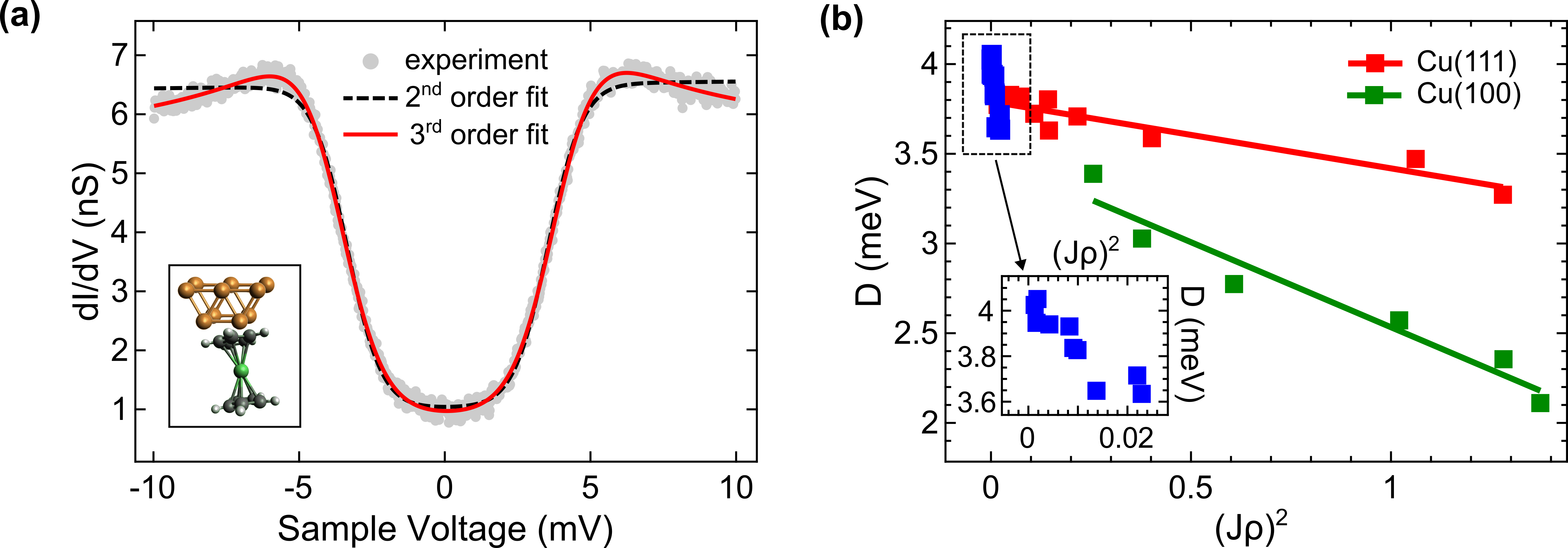}
\caption{Third-order analysis for tunneling spectra acquired with different Nc tips.
(a) Spectrum of $\text{d}I/\text{d}V$ (dots) of an Nc-terminated tip above Cu(111) (feedback loop parameters: $-10\,\text{mV}$, $50\,\text{pA}$) together with second-order (dashed line) and third-order (solid line) fits.
(b) Comparison of $D$-versus-$(J\varrho)^2$ data (squares) extracted from third-order fits to $\text{d}I/\text{d}V$ spin excitation spectra of 10 Nc-terminated tips (blue) with results obtained from Nc tips approaching Cu(111) (red) and Cu(100) (green)\@.
Solid lines are linear fits to the data. 
Inset: close-up view of (b) for data obtained with the ten tips in the far tunneling range.}
\label{figS6}
\end{figure}

Figure 2d of the article presents the collection of spin excitation energies $D$ obtained from ten different Nc-terminated tips in the far tunneling range above pristine Cu(111) (feedback loop parameters for spectroscopy: $10\,\text{mV}$, $30\,\text{pA}$--$100\,\text{pA}$)\@.
The $\text{d}I/\text{d}V$ spectrum of the lower inset to Fig.\,2d is repeated here (Fig.\,\ref{figS6}a)\@.
Superimposed with the experimental data (dots) are fits within the second-order (dashed line) and third-order (solid line) scattering model (see Sec.\,S5)\@.
Obviously, the latter provides an improved match compared with the former because the cusps at $\approx\pm 6\,\text{mV}$ are well reproduced for $J\varrho = -0.15\pm0.01$.
Indeed, the reliability factors $R^2$ of the least-mean-squares fits are $R^2=0.995$ (second order) and $R^2=0.998$ (third order)\@.
Importantly, the third-order contributions ($\propto |J\varrho|$, eq \ref{eq:sigma3}) signaled by cusps are very weak over all tested Nc tips (Fig.\,\ref{figS6}b) and thus, the extraction of $D$ via the second-order fit to tunneling spectroscopic data as done for Fig. 2d of the article is justified.

Because of the applicability of the third-order model to tunneling spectra of Nc tips it is tempting to assume Kondo exchange coupling of Nc with the tip it is attached to as rationale for their wide $D$ variation.
To further explore this assumption, the $D$-versus-$(J\varrho)^2$ behavior was extracted for all tips of Fig.\,2d via third-order fits.
Since, to first order, the Nc--tip coupling is conceptually similar to approaching an Nc tip to Cu(111) or Cu(100), a similar D variation is expected.
In fact, fig. \ref{figS6}b (blue squares) shows a strong variation of $D$ across the 10 tips for nearly vanishing $J\varrho$.
A linear fit to these data yields a slope of $-17\,\text{meV}$, which exceeds the slope observed for approaching an Nc tip to Cu(111) (red squares) and Cu(100) (green squares) by more than an order of magnitude (see sec. S6)\@.
Therefore, the scattering of $D$ is most likely not driven by the Kondo exchange interaction.
Rather, it is surmised that Kondo exchange coupling is dominated by Nc--tip charge transfer \cite{nl_17_1877}, which may depend on the actual tip geometry.

\providecommand{\latin}[1]{#1}
\makeatletter
\providecommand{\doi}
  {\begingroup\let\do\@makeother\dospecials
  \catcode`\{=1 \catcode`\}=2 \doi@aux}
\providecommand{\doi@aux}[1]{\endgroup\texttt{#1}}
\makeatother
\providecommand*\mcitethebibliography{\thebibliography}
\csname @ifundefined\endcsname{endmcitethebibliography}
  {\let\endmcitethebibliography\endthebibliography}{}